\documentstyle[twocolumn,prl,aps,psfig]{revtex}

\begin{document}
\title{Novel Collective Excitation in Spin Textured Edges 
of  Quantum Hall Systems.}
\author{M. Franco and L.Brey}
\address{Instituto de Ciencia de Materiales (CSIC),
Cantoblanco,  28049 Madrid, Spain}
\date{\today}
\maketitle

\begin{abstract}
We study the electric and magnetic properties of
the edge of a two-dimensional electron gas in presence of a magnetic
field and at filling
factor unity. 
The existence of a spin textured
edge is proved as a function  of the Zeeman coupling and of the smoothness
of the confining potential.
We also calculate the low energy excitation of the spin textured phase.
We obtain that in addition to the classical edge magnetoplasmons at small
wavectors, there is an 
almost dispersionless  excitation, 
with  a finite gap of energy  at zero wavevector.
This excitation is  associated with the broken symmetry of the
spin textured edge phase.

\end{abstract}


There is great interest in  understanding  the properties
of the edge states of a two-dimensional electron gas (2DEG)
in presence of a strong magnetic field $B$ in the 
quantum Hall effect (QHE) regime. 
The edge states are important because they can  control the 
magnetotransport of
the 2DEG in a broad class of mesoscopic and macroscopic systems.\cite{book1}
Also, under normal conditions the only gapless excitations in the QHE regime
are edge excitations.\cite{wen}

Because of the screening properties of the 2DEG, 
the structure of the edge states changes when the 
smoothness of the edge confining potential, $V_0 (x)$, varies.  
For studying the edge states properties,  we consider a semi-infinite $x-y$ 
plane, with a straight
edge parallel to the ${\bf \hat y}$ direction. 
In the case of sharp confinement potentials  the electron density falls from the 
bulk value to zero in a distance of the order of the 
magnetic length $\ell$=$\sqrt{\hbar c / e B}$.
For a sufficiently smooth confining potential, 
it has been proposed theoretically that 
the edge separates
into incompressible and compressible regions.\cite{smooth1,smooth2}
This picture appears to be in reasonable agreement 
with recent experiments.\cite{exp1}
For intermediate smooth 
confining potentials, 
due to the  exchange interactions
it has been suggested that 
stripes of charge density 
corresponding to filling factor $\nu$=1 can be stabilized at the
edge of the 2DEG.\cite{herb1,chamon}
This edge reconstruction can explain  transport experiments in
quantum dots\cite{klein}.
When the spin degree of freedom is included in the theoretical models,
new effects such as 
spontaneous spin polarization\cite{dempsey}  
and spin textures\cite{karlhede,oaknin,lubin} at the edges of quantum Hall systems 
have  been predicted.

The low-energy collective excitations localized at the edge of the system
have  dispersion relations  that
depend strongly  on the type of reconstruction
at the edge.
For a sharp edge, a magnetoplasmon with a dispersion 
of the form 
$q \ln {q}$   is expected.\cite{volkov,prepexp} Here 
$q \parallel {\bf \hat y}$  is the wavevector of the excitation.
For smooth compressible edges, 
in addition to the magnetoplasmon,
the existence of new branches of acoustic excitations 
have been proposed\cite{aleiner} 
and measured\cite{ernst}. 
For the striped phase, the collective excitations consist of  a
set of $q \ln {q} $ branches and a number of  acoustic modes equal
to the number of stripes present at the edge of the 2DEG.\cite{unp}

In this work we study the electrical and magnetical properties of the
edge states of the 2DEG at $\nu$=1 as a function of the Zeeman
coupling,  $\tilde g = g \mu _B B$, and of the smoothness of the confining
potential. 
In particular we are interested in the existence and properties
of a spin textured edge.
Edge textures are configurations of the spin density that possess
a topological charge density at the edge of the system.
The spin field, ${\bf n} ( {\bf r})$, 
of the spin texture has the form\cite{karlhede}
\begin{equation}
n_x + i n _y \! = \!  \sqrt{1 \!  - \! f ^2 (x)} \, \, e ^ {i (G_s y + \theta)} \, \, \, 
, \, \, \,  
n_z=f(x)\, \, \, ,
\end{equation}
here $G_s$ is the wavevector of the spin texture and $\theta$
is an arbitrary phase. In the polarized bulk we have
$f(x)=-1$. 
In the QHE regime the topological charge density coincides 
with  the real charge density\cite{sondhi,hf}.
The charge density associated with the texture only depends 
on $x$ and is proportional to $G_s   d f / d x$.        
The system can develop a spin texture 
in order to modulate 
the charge density profile in the ${\bf \hat x}$ direction
and therefore  to screen the edge confinement.
The lowering of the confinement energy competes with the 
cost in exchange and Zeeman energies. Therefore the spin textured edge 
occurs only for smooth enough confinement potential and for small enough
Zeeman coupling.
\par \noindent
The two main results of this  work are the following:
\par \noindent 
i) We obtain the  phase diagram of the system as a function of the
smoothness of the confining potential and of the 
Zeeman energy.
By performing a full unrestricted Hartree-Fock (HF) calculation we obtain 
the range of parameters where a spin textured edge can be expected.
Given a confining potential, we obtain that the maximum value of $\tilde 
g$ where the
spin texture exists is considerable 
smaller than the obtained in reference [10].
This discrepancy occurs because reference [10]
only considers the competition of 
the spin textured state  with the
striped  phase. 
However, we obtain that  making $V_ 0$ smoother,
the sharp edge becomes
unstable to  a smooth charge modulation in both 
the ${\bf x}$ and ${\bf y}$ directions
before  it becomes 
unstable to  the stripe phase.
\par \noindent
ii) We have studied  the dispersion relation of the collective excitations
in a spin textured edge phase.
We  obtain that in addition to magnetoplasmon like excitations 
and the bulk spin density waves, there 
exists a low energy excitation associated with the broken symmetry
of this phase.  
This mode is almost dispersionless, and it has a finite energy 
at zero $q$. The existence of this gap is due to the finete width
of the spin texture in the ${\bf \hat x }$ direction.\cite{internal}
We turn now to the details of  our calculations and results.

{\it Microscopic Hamiltonian and HF approximation.-}
We are interested in properties of the edge states of the $\nu$=1
quantum Hall state. In this regime we assume that 
the electron-electron interaction and Zeeman energies are  much smaller than the
Landau level splitting, and 
we therefore 
restrict the orbital Hilbert space to the lowest Landau level. 
Since the confining potential only depends on the $x$ coordinate, it is 
convenient to work in the Landau gauge, ${\bf A} = B x {\bf \hat y } $.
The Hamiltonian of this system has the form (Through this work we take 
$\ell$ as the unit of length and $e^2/\ell \epsilon$ as the
unit of energy):
\begin{eqnarray}
H=  & & \sum   _ {k, \alpha}  \left ( V_0 ( k) + { \tilde g \over 2 } 
\alpha \right ) c _{k, \alpha }^ {\dagger}
c _{k, \alpha }   \nonumber \\
 \, \, \, \, \, \, \, \, \, \,  \, \, \, \, \, \,+ &  &  { 1 \over {2 L_x L _y }} 
\sum _{k, k ^{\prime}  ,\vec{p},\alpha , \beta}
 v(p) e^ {-p^2/2} e^{i p_x ( k- k^{\prime} + p_y)} \nonumber \\
& & \, \, \, \, \, \, \, \, 
\,\, \, \, \, \, \, \, \, 
\,\, \, \, \, \, \, \, \, 
\,\, \, \, \, \, \, \, \, 
\,\, \, \, \, \, \, \, \, 
\,
c _{k, \alpha }^ {\dagger}
c _{k ^{\prime} , \beta}^ {\dagger}
c _{k ^{\prime} + p_y  , \beta}
c _{k  - p_y  , \alpha} \, \, .  
\end{eqnarray}
Here $\alpha$, $\beta$=+(up)-(down) are 
spin indices,
$L_x$ and $L_y$ are the sample dimensions, $v(p)$ is the Fourier
transform of the Coulomb interaction, 
and
$c _{k, \alpha }^ {\dagger}$ creates an electron with spatial  wavefunction 
$\psi _ k ({\bf r})= {1 \over { \sqrt{L_y \pi ^{1/2}} } }
e^{i k y } e ^{- (x-k)^2 / 2}$, and spin  $\alpha / 2$, 

In order to change  the smoothness of the edge continuosly we take 
$V_0$   as the potential created by a distribution of positive charge
which falls linearly from its bulk value,  $ 1/ 2 \pi$,  to zero, over a region
of width $W$.
In this way the edge is smoother in direct proportion to  $W$.
This form of $V_0$ has been used in previous 
works\cite{smooth1,smooth2,chamon,karlhede}. For small values of
$W$ the edge is sharp. Chamon {\it et al}\cite{chamon}, 
obtained that in the HF approximation
the stripe phase is stable with respect the sharp edge
for values of $W$ larger  than  $W_c$=9.0.
Working also in the HF approximation
Karlhede {\it et al}\cite{karlhede}
obtained a critical value of $W_s$=6.8 for the existence  of a textured edge
at $\tilde g$=0.
These two calculations do not allow the  charge density to vary 
along the edge.
In this work we allow the system to modulate the charge in both directions
${\bf \hat x}$ and ${\bf \hat y}$. 

In order to solve the Hamiltonian Eq.1 we make the  HF
pairing of the second-quantized operators, allowing for the
possibility of different  broken translational symmetry and spin 
order in the ground state.
To characterize  the different solutions, it  is very convenient
to introduce the operators
\begin{equation}
\rho _ {\alpha ,\beta }({\bf q})= 
{2 \pi \over { L_x L_y}}  \sum _{k} e^{ -iq_x (x -  q_y /2)}
c _{k, \alpha }^ {\dagger} 
c _{k+q_y , \beta} 
\end{equation}
which are easily related to the charge  $ n({ \bf  q})$  and spin ${\bf S } ({\bf q})$
density operators.\cite{cote}
By solving selfconsistently the Hartree-Fock equations we obtain 
the espectation
values
of the energy and of the different density operators.

{\it Phase Diagram.-} The different solutions of the electric and magnetic
edge structure can be 
characterized by the expectation values of the 
products 
$c _{k, \alpha }^ {\dagger} c _{k ^ {\prime} , \beta} $ or by the expectation 
values of the operators 
$ \rho _ {\alpha ,\beta }({\bf q})$.
In this work we find the following type of solutions:
(see Fig 1):
\par 
i) Spin polarized compact edge (SPCE). In this state 
$<\! \! c _{k, \alpha }^ {\dagger} c _{k ^ {\prime} , \beta} \! \!>$ 
=$ \delta _{k,k ^ \prime} \delta _{\alpha,\beta} \delta_{\alpha,-}$, and there
is a maximum  wavevector such that all states with smaller momentum 
are occupied.
This solution is the sharpest edge possible, and it is the ground state
for small values of $W$.
\par 
ii) Spin polarized charge density wave (SPCDW). 
In this state 
only the majority spin electronic states are occupied, i.e.
$<\! \!  \rho _ {\alpha ,\beta }({\bf q})\! \!> \propto
 \delta _{\alpha,\beta} \delta_{\alpha,-}$. 
In this class of solutions, 
the system  modulates the charge along the
${\bf \hat x }$ direction,
in order to screen the edge
potential. 
In the QHE regime the system only can modulate 
smoothly the charge along the ${\bf \hat x}$ direction by modulating also the charge along
the ${\bf \hat y  }$ direction.
At $\tilde g \rightarrow \infty $,  
the SPCDW  state has lower  energy than the SPCE state for
$ W>W_{CDW} \simeq 7$,  while 
the stripe phase has lower energy than the SPCE at $W_c=9.0$\cite{chamon}. 
We find that the SPCDW state  always has lower energy than the stripe phase: 
the stripe phase is not a stable solution of the system.
\par 
iii) Spin textured edge (STE). In this class of solutions, 
$<\! \! \rho _ {\alpha ,\alpha  }({\bf q}) \!\!> \propto \delta_{q_y,0}$ but all
the 
$<\!  \!\rho _ {\alpha ,- \alpha  }({\bf q}) \!\!>$ can be different from zero.
In the calculation we obtain that the operators
$\rho _{\alpha, -\alpha} ( \bf q)$ are different from zero only for one  
wavevector of the form 
${\bf q }=(0,G_s) $. Minimizing the energy with respect ito $G_s$ we obtain 
microscopically the
periodicity of the spin texture.
We do not obtain higher harmonics of the spin texture because,
in order to get a constant charge density along the edge, 
{\it only } one   
wavevector of the spin texture is possible.\cite{fieldth}

Since 
$<\!  \!\rho _ {\alpha ,- \alpha  }(0,G_s)\! \!>$ is  different from zero,
the STE phase breaks  
the translational invariance along the edge and the
spin rotational symmetry about the magnetic field. However the STE is invariant
under a symmetry composed of a translation along the edge and a 
spin rotation.\cite{karlhede}
The states related with this symmetry correspond to the different values
of $\theta $ in Eq.1

For  $\tilde g=0$, the STE  has lower energy than the 
SPCE  for $W> W_s=6.7 $. 
This value of $W_s$ increases with $\tilde g$, and for 
$\tilde g >  \tilde g _c \simeq  0.008$ the system prefers to screen the edge potential
by forming a SPCDW state rather than by creating a STE.
This value of $\tilde g _c$ is about ten times smaller than the obtained  by
Karlhede {\it et al}\cite{karlhede}. This  discrepancy occurs
because in reference [10] the STE is assumed to compete only with the striped phase,
and not with the SPCDW state.

It is important to note that the width  of the charge and spin 
modulation in the ${\bf \hat x}$ 
direction can be much larger than $G_s$. 
For example for $\tilde g$=0 and
$W$=8, $G_s \approx 0.85$ and the width of the modulation is of the order of
4 magnetic lengths.

\par 
iv) Spin textured and charge density wave state. This is a fully 
broken symmetry ground state where  the expectation value of all
$<\! \! \rho _ {\alpha ,\beta }({\bf q})\! \!> $ are not  zero. 
This state is a mixture of charge density waves and spin textures and
it  is reached from both the STE and SPCDW states 
by making the edge confinement smoother. This phase corresponds to the shadow region 
in Fig.1

{\it Collective excitations.-}
We now study the collective excitations of the spin textured edge. As
commented above, in the STE phase the only order parameters different from zero are:
$<\! \! \rho _ {\alpha , -\alpha  }(0,G_s )\! \!>$ and 
$<\!  \!\rho _ {\alpha , \alpha  }(q_x,0)\! \!>$.
With these order parameters the  wavefunction of the STE is a Slater determinant
of the form $|STE \! >$ =$\prod _{k } d ^{ \dagger} _k |0 \!  \!>$, where
$d ^ {\dagger} _k = u_k c ^{\dagger} _{-, k} 
+ v_k c^{\dagger} _ {+,k+G_s}$, 
and   $u_k$ and $v_k$  are obtained from 
the expectation values of the 
$\rho _ {\alpha , \beta  }$ operators.
and verify the relation $ u_k ^2 + v _k ^ 2 $=1.
The set of  eigenstates of the Hartree-Fock Hamiltonian orthogonal
to the $d^{\dagger}$ states have the form:
$b ^ {\dagger} _k = - v_k c ^{\dagger} _{-, k} 
+ u_k c^{\dagger} _ {+,k+G_s}$. 

Using the $d$ and $b$ states, the low energy collective excitations of the
system can be characterized by a quantum number $q$ and correspond
to linear combinations of electron-hole pairs of the
form
$d^{\dagger} _{k+q} d _q $ and $ b ^{\dagger} _{k+q}  d_k  $.
From the form of the $d$ and $b$ operators the excitations have the 
general expression
\begin{equation}
|\psi _q ^ i >=  \! \sum _{k,\alpha,\alpha ^{\prime}} 
a ^ i _{ k,\alpha, \alpha ^{\prime}} (q) \, \, 
c_{k+g_{\alpha}+q, \alpha} ^{\dagger}
 c_{k+g_{\alpha ^\prime }, \alpha ^\prime } 
|STE \! > \, \, \, \, \,  , 
\end{equation}
where $g_{-}$=$0$ and $g_{+}$=$G_s$, and 
the coefficients 
$a ^ i _{ k,\alpha, \alpha ^{\prime}}$
are obtained by minimizing the 
energy of the excitations, $\hbar \omega ^i (q) $=$ < \!  \psi ^i _q |H|\psi^i _q  \! >
\! \! /\! \!  <\!  \psi ^i _q |\psi^i _q \!  >$. 

Due to the existence of the spin texture, the collective excitations are a 
mixture of spin and charge density excitations.
Also note that because the STE is invariant under a translation along the 
edge plus a spin rotation\cite{karlhede}, any electron spin flip is
accompanied  by a change of the electron wavevetor in  $\pm G_s$. 

In Fig. 2 we plot the lowest energy collective excitations for the
$\tilde g$=0 and $W$=8 case. 
The expectrum consists basically of two parts:
i) a continuum of excitations  
starting from an  energy  $ \tilde {g} + 4 \pi \rho _s (q-G_s)^2$, where 
$\rho_s$ is the spin stiffness of the 2DEG at $\nu$=1, and 
ii) a set of discrete branches around $q$=0. 
The former excitations are localized in the bulk part of the system and they
are the well known bulk spin density waves.\cite{bychkov} 
The dispersion relation of the spin density waves starts at $q=-G_s$, because
in the STE phase the electron hole pairs involving 
a majority spin flip have the form (see Eq. 4) 
$c ^ {\dagger} _{k+G_s+q, +}
 c_{k, - } $.
On the other hand  the  low energy excitations starting at $q$=0 are localized
at the edge of the system and they correspond to  edge 
excitations of the STE phase.

We describe now the character of  the  edge excitations of the STE phase. 
At small wavevectors, all but one of  the low  energy excitations 
are gapless at $q$=0, and 
have
a dispersion of the form $q \ln q$.
The analysis of the coefficcients  
$a ^ i _{ k,\alpha, \alpha ^{\prime}}$ 
of these gapless excitations, reveals that
they are localized at  the edge of the
system in a region of thickness $q$. These  excitations correspond to
the classical
edge magnetoplasmons\cite{volkov,nosotros}
of the system. The difference with the edge magnetoplasmon of the 
spin polarized compact edge is that in the STE  the spin and charge
excitations  are mixed.

As mentioned above, in adition to the magnetoplasmon, there is a low energy excitation which is 
practically dispersionless at small wavevectors and which has a finite gap at 
$q$=0. 
This excitation anticrosses with the edge magnetoplasmon, see inset of Fig.2.
It is localized at the edge of the system but 
with a thickness equal to the spatial width  of the charge modulation
in the ${\bf \hat x }$ direction.    
This thickness is much bigger than the wavevector of the excitation,
and therefore  this mode is almost dispersionless in $q$.
The  coefficcients  
$a ^ i _{ k,\alpha, \alpha ^{\prime}}$ 
corresponding to this mode  
show that this excitation is one in which the transverse component 
of the spin polarization becomes time dependent. Also there is a
motion of the charge density across the edge. 
This mode is related with the broken symmetry occuring in the STE phase.
In a classical calculation this mode should be gapless at $q$=0, but a finite 
gap it is expected\cite{internal} in quantum calculations because of the finite
size of the STE.
The energy of this mode at $q$=0, is around 0.01 and  is not very dependent on the
value of 
Zeeman energy nor  on the confinement potential.
Because of the external potential 
at positive large values of $q$, 
the energy of this excitation increases. 
For negative values of $q$ we find
that this excitation is damped into the spin density wave region
at wavevectors of the order of $-G_s$.

{\it Experimental consequences.}
From the phase diagram Fig.1 the maximum value of the Zeeman energy 
for the existence of the STE phase is $\tilde g$=0.008
This is a rather small value but it can be reached in  GaAs systems
by applying hydrostatic pressure.\cite{maude}
In order to get a  STE phase 
a smooth confining
potential
is also necessary.
It is possible to tune the edge  potential to the appropriate value,
by applying gate bias to the edge of the 2DEG.\cite{zhitenev1}

The existence of the STE affects
the  spin polarization  of the 2DEG edge at $\nu$=1.
By using a local NMR probe\cite{explo} 
the polarization
of an edge as a function of the Zeeman coupling or of  the
strength of the confinement potential
can be studied. 
As in the case of
Skyrmions\cite{hf} a variation
of the spin polarization with these parameters should probe the existence
of the STE phase.\cite{barrett} 

As discussed above,
in the  STE phase there exists 
a low energy excitation with a finite gap at $q$=0.
The detection of this mode propagating along the edge of a 2DEG
should be a probe of the existence of the STE phase. 
It could be possible to detect the existence of this mode by 
time-resolved magnetotransport experiments\cite{exp1}
or by measuring the transmission of electromagnetic waves.\cite{prepexp}

In summary, we have studied the electronic and magnetic structure
of the edge of a 2DEG in the $\nu$=1 QHE regime.
We have obtained the phase diagram of the system as a function
of the Zeeman coupling and the smoothness of the confinement potential.
We obtain the range of parameters where a spin textured edge phase
is expected. We have also studied the collective excitations of this
phase. We have found the  existence of a low energy gapfull collective excitation 
associated with the broken symmetry of the spin textured phase.

This work was supported by 
CICyT of Spain under contract MAT 94-0982.
Helpful conversations with 
Luis Martin-Moreno, J.P.Rodriguez   and 
Carlos Tejedor are gratefully
acknowledged.

\begin{figure}
\caption{ }
Phase diagram, as a function of $\tilde g$= $ g \mu _B B $  and $W$, of the edge of a 2DEG at  
$\nu=1$. 
The shadow region corresponds to the spin textured and charge density wave phase.
$\tilde g$ is in units of $e^2 /\epsilon \ell$ and $W$ in units of $\ell$.
\end{figure}

\begin{figure}
\caption{}
Low energy collective excitations
of the spin textured edge phase. The results corrrespond to
$\tilde g$= $ g \mu _B B $=0.   and $W$=8. 
The energy is in units of $e^2 /  \epsilon \ell$ and the wavevector $q$ in units of $\ell ^{-1}$.

\end{figure}


\begin{references}

\bibitem{book1}C.W.Beenakker and H. van Houten, in {\it Solid State Physics},
edited by H.Ehrenreich and D.Turnbull (Academic, NY, 1991), Vol.44, p.1,
and references therein.
\bibitem{wen}X-G. Wen, Phys.Rev.B {\bf 41}, 12838 (1990).
\bibitem{smooth1} C.W.L.Beenakker, Phys.Rev.Lett. {\bf 64}, 220 (1990);
A.M.Chang, Solid State Comm. {\bf 74}, 871 (1990); 
L.Brey, Phys.Rev.B {\bf 50}, 11861 (1994).
\bibitem{smooth2}D.B.Chklovskii {\it et al.} Phys.Rev.B {\bf 46}, 4026 (1992);
L.Brey {\it et al} Phys.Rev.B {\bf 47}, 13884 (1993).
\bibitem{exp1}N.B.Zhitenev {\it et al.} Phys.Rev.Lett. {\bf 71}, 2292 (1993);
S.Takaoka {\it et al, ibid} {\bf 72}, 3080 (1994).
\bibitem{herb1}H.A.Fertig {\it et al.} Phys.Rev.B, {\bf 47}, 10484 (1993).
\bibitem{chamon}C. de C. Chamon {\it et al} Phys.Rev. B {\bf 49}, 8227 (1994).
\bibitem{klein}O.Klein {\it et al.} Phys.Rev.Lett. {\bf 74}, 785 (1995).
\bibitem{dempsey}J.Dempsey {\it et al.} Phys.Rev.Lett. {\bf 70}, 3639 (1993).
\bibitem{karlhede}A.Karlhede {\it et al.} Phys.Rev.Lett. {\bf 77}, 2061 (1996).
\bibitem{oaknin}J.H.Oaknin {\it et al.} Phys.Rev.B {\bf 54}, 16850 (1996).
\bibitem{lubin}M.I.Lubin {\it et al.} Cond-Matt/9701079.
\bibitem{volkov}V.A.Volkov and S.A.Mikhaelov, Zh. Eksp. Teor. Fiz.
{\bf 94}, 217 (1988) [Sov.Phys. JETP {\bf 67}, 1639 (1988)].
\bibitem{prepexp}N.Q.Balaban {\it et al.} Cond-Matt/9702110.
\bibitem{aleiner}I.L.Aleiner {\it et al.}, Phys.Rev.Lett. {\bf 72}, 2935
(1994); Phys.Rev.B {\bf 51}, 13467 (1995).
\bibitem{ernst}G.Ernst {\it et al.} Phys.Rev.Lett. {\bf 77}, 424 (1996).
\bibitem{unp}M.Franco and L.Brey (unpublished).
\bibitem{sondhi}  S.L. Sondhi, {\it at al.} Phys. Rev. B {\bf 47}, 16419 (1993).
\bibitem{hf}  H.A. Fertig, {\it et al.} Phys. Rev. B {\bf 50}, 11018 (1994). 
\bibitem{internal} H.A.Fertig, {\it et al.},
Phys.Rev.Lett. {\bf 77}, 1572 (1996).
\bibitem{cote}R.C\^{o}t\'{e} {\it et al.}, Phys.Rev.B {\bf 44}, 8759 (1991); {\it ibid}
{\bf 51}, 13475 (1995); {\it ibid} {\bf 53}, 10019 (1996).
\bibitem{fieldth}Using the relation between spin texture 
and charge density\cite{sondhi},
it is easy to prove that spin textures with a dependence on the coordenate $y$ 
different  
than a  $sine $  implies 
charge modulation in the ${\bf \hat y}$ direction. 
\bibitem{bychkov}Y.U.Bychkov {\it et al.}, Pis´ma Zh. Eksp. Teor. Fiz. {\bf 33}, 152 (1981)  
[JETP Lett. {\bf 33}, 143 (1981)].
\bibitem{nosotros}M.Franco and L.Brey, Phys.Rev.Lett. {\bf 77}, 1358 (1996).
\bibitem{maude}D.K.Maude {\it et al.} Phys.Rev.Lett. {\bf 77}, 4604 (1996).
\bibitem{zhitenev1} N.B.Zhitenev {\it et al.} Phys.Rev.Lett. {\bf 77}, 1833 (1996).
\bibitem{explo} K.R.Wald {\it et al.}, Phys.Rev.Lett. {\bf 73}, 1011 (1994).
\bibitem{barrett}  R. Tycko {\it et al.}
Science {\bf 268}, 1460 (1995); S.E. Barrett {\it et al.}
Phys.\ Rev.\ Lett.\ {\bf 74}, 5112 (1995).

\end{references}
\end{document}